\documentclass[sigconf, nonacm]{acmart}

\AtBeginDocument{%
  }

\copyrightyear{2025}
\acmYear{2025}
\setcopyright{cc}
\setcctype{by}
\acmConference[WWW '25]{Proceedings of the ACM Web Conference 2025}{April 28-May 2, 2025}{Sydney, NSW, Australia}
\acmBooktitle{Proceedings of the ACM Web Conference 2025 (WWW '25), April 28-May 2, 2025, Sydney, NSW, Australia}
\acmDOI{10.1145/3696410.3714790}
\acmISBN{979-8-4007-1274-6/25/04}

\begin{document}

\title[SuiGPT MAD: Move AI Decompiler to Improve Transparency and Auditability \\ on Non-Open-Source Blockchain Smart Contract]{SuiGPT MAD: Move AI Decompiler to Improve Transparency and Auditability on Non-Open-Source Blockchain Smart Contract}

\author{Eason Chen}
\email{eason.tw.chen@gmail.com}
\additionalaffiliation{%
  \institution{Mysten Labs}
  \city{Palo Alto}
  \state{CA}
  \country{USA}
}
\affiliation{%
  \institution{Carnegie Mellon University}
  \city{Pittsburgh}
  \state{PA}
  \country{USA}
}
\orcid{0000-0003-1486-8559}

\author{Xinyi Tang}
\affiliation{%
  \institution{Carnegie Mellon University}
  \city{Pittsburgh}
  \state{PA}
  \country{USA}
}
\orcid{0009-0001-1089-8550}

\author{Zimo Xiao}
\affiliation{%
  \institution{Carnegie Mellon University}
  \city{Pittsburgh}
  \state{PA}
  \country{USA}
}
\orcid{0000-0002-9473-6056}

\author{Chuangji Li}
\affiliation{%
  \institution{Carnegie Mellon University}
  \city{Pittsburgh}
  \state{PA}
  \country{USA}
}
\orcid{0009-0000-5408-0761}

\author{Shizhuo Li}
\affiliation{%
  \institution{Carnegie Mellon University}
  \city{Pittsburgh}
  \state{PA}
  \country{USA}
}
\orcid{0009-0008-2910-5451}

\author{Tingguan Wu}
\affiliation{%
  \institution{Carnegie Mellon University}
  \city{Pittsburgh}
  \state{PA}
  \country{USA}
}
\orcid{0009-0005-0809-1482}

\author{Siyun Wang}
\affiliation{%
  \institution{Carnegie Mellon University}
  \city{Pittsburgh}
  \state{PA}
  \country{USA}
}
\orcid{0009-0005-8017-7475}

\author{Kostas Kryptos Chalkias}
\affiliation{%
  \institution{Mysten Labs}
  \city{Palo Alto}
  \state{CA}
  \country{USA}
}
\orcid{0000-0002-3252-9975}


\renewcommand{\shortauthors}{Eason Chen et al.}

\begin{abstract}
The vision of Web3 is to improve user control over data and assets, but one challenge that complicates this vision is the prevalence of non-transparent, scam-prone applications and vulnerable smart contracts that put Web3 users at risk. While code audits are one solution to this problem, the lack of smart contracts source code on many blockchain platforms, such as Sui, hinders the ease of auditing. A promising approach to this issue is the use of a decompiler to reverse-engineer smart contract bytecode. However, existing decompilers for Sui produce code that is difficult to understand and cannot be directly recompiled. To address this, we developed the SuiGPT Move AI Decompiler (MAD), a Large Language Model (LLM)-powered web application that decompiles smart contract bytecodes on Sui into logically correct, human-readable, and re-compilable source code with prompt engineering. 

Our evaluation shows that MAD's output successfully passes original unit tests and achieves a 73.33\% recompilation success rate on real-world smart contracts. Additionally, newer models tend to deliver improved performance, suggesting that MAD's approach will become increasingly effective as LLMs continue to advance.

In a user study involving 12 developers, we found that MAD significantly reduced the auditing workload compared to using traditional decompilers. 
Participants found MAD’s outputs comparable to the original source code, improving accessibility for understanding and auditing non-open-source smart contracts.
Through qualitative interviews with these developers and Web3 projects, we further discussed the strengths and concerns of MAD.

MAD has practical implications for blockchain smart contract transparency, auditing, and education. It empowers users to easily and independently review and audit non-open-source smart contracts, fostering accountability and decentralization. Moreover, MAD's methodology could potentially extend to other smart contract languages, like Solidity, further enhancing Web3 transparency.


\end{abstract}





\maketitle

\section{Introduction}

Web3, also referred to as the decentralized web, represents a significant shift in the evolution of the internet \cite{nath2014web, ray2023Web3}. Unlike earlier versions—Web1, which focused on static content, and Web2, which emphasized user-generated content and platforms—Web3 introduces decentralization, prioritizing user autonomy and ownership through blockchain technology \cite{zheng2018blockchain}. A central element of Web3 is the use of smart contracts \cite{vacca2021systematic}, which are self-executing, algorithmic-based agreements that automatically enforce terms. Smart contracts enable automated, trustless transactions, fostering the decentralized nature of Web3 ecosystems. This transition marks a major restructuring of internet architecture, shifting control from centralized entities to individuals and decentralized networks \cite{zetzsche2020decentralized}.

Nevertheless, the increased autonomy and control that Web3 grants to its users also demands that they take on greater accountability for their decisions within this decentralized ecosystem. Web3 has witnessed significant instances of fraud and vulnerabilities \cite{chen2020survey, vacca2021systematic}, leading to substantial financial losses \cite{Over1bni65:online}. Notably, in the first half of 2024, over \$1 billion in assets were lost in Web3-related cryptocurrencies, with phishing attacks (n = 150) and smart contract vulnerabilities (n = 105) being the most frequent causes \cite{Over1bni65:online}. In the absence of centralized regulators or platforms to verify contract security, one viable path is to empower users to audit smart contracts independently. By enabling users and communities to perform these audits, Web3 not only strengthens the transparency and accountability of decentralized applications but also reduces reliance on third-party authorities for security verification \cite{aggarwal2019blockchain}. Independent auditing ensures that the decentralized ethos is upheld while mitigating risks inherent in open, trustless environments.


Unfortunately, on emerging blockchain platforms like Sui \cite{TheSuiSm42:online}, smart contracts are published as bytecode, and their source code is not always open-source. This creates several potential risks, for example, the following are some real-world case studies:

\begin{enumerate}
    \item Scammers create meme coins that \textbf{block buyers from selling at the smart contract logic}, causing the coin price to keep rising until scammers sell all their holdings.
    \item Developers create \textbf{unfair games} where certain actions have a higher chance of winning, leading to exploitation.
    \item Decentralized exchanges contain \textbf{unexpected backdoors}, allowing users to withdraw assets deposited by other users.
\end{enumerate}

If the smart contract source code were accessible, the aforementioned risks could be easily identified. However, these contracts are not open-source, and even if the source code is available on platforms like GitHub, users cannot be certain that it is the contract deployed on the blockchain. This underscores the critical need for \textbf{transparency} in smart contracts within the Web3 space. There is a need for a \textbf{effective, easy-to-use application} that allows users to independently \textbf{audit} the logic of non-open source smart contracts, improve \textbf{transparency}, and promote \textbf{algorithmic accountability} in the Web3 ecosystem. Our paper seeks to address this gap.

\textbf{Our study} developed and evaluated the SuiGPT Move AI Decompiler (MAD). MAD is a Large Language Model (LLM) powered web application that converts Move smart contract bytecode into easily readable and re-compilable source code, enabling developers to understand and audit non-open-source smart contracts on Sui.
We summarized our main contribution as follows:


\begin{enumerate}
    \item We developed and evaluated SuiGPT MAD, an AI-powered web application that generates logically correct, human-readable, and re-compilable Move code from bytecode, achieving a 73.3\% recompilation success rate and facilitating transparency and auditability in non-open-source smart contracts.
    \item We conducted a user study involving 12 developers, revealing that MAD significantly reduced code comprehension workload compared to existing decompilers, with its output perceived as comparable to the original source code.
    \item Through interviews, we explored MAD’s practical applications in enhancing Web3 transparency, auditing, and education, along with the challenges and concerns it may pose.
\end{enumerate}

\section{Background and Related Works}

\subsection{The Move Programming Language}
In the current blockchain ecosystem, Solidity \cite{wood2014ethereum} remains the predominant language for smart contract development \cite{ray2023Web3}. However, it is also prone to vulnerabilities such as reentrancy attacks and integer overflows \cite{zou2019smart}, which have been exploited by malicious actors, leading to significant financial losses \cite{chen2020survey, vacca2021systematic, Over1bni65:online}. In response to these security concerns, several blockchain platforms have chosen to use Move, a new domain-specific language aimed at addressing the shortcomings of Solidity. For instance, emerging blockchains like Sui \cite{TheSuiSm42:online}, Aptos \cite{AptosWhi99:online}, Diem \cite{TheLibra87:online}, IOTA \cite{iotaIOTARebased}, and Movement \cite{movement9:online} have adopted the Move language \cite{blackshear2019move} for smart contract development.

Move language \cite{blackshear2019move}, originally developed for Meta's Diem (formerly Libra) project \cite{TheLibra87:online}, was designed to address many vulnerabilities found in Solidity by incorporating several key features. Specifically, Move offers the following two key advantages to enhance security and prevent common vulnerabilities:

\begin{enumerate}
    \item \textbf{Resource Management and Safety}: Unlike Solidity, which lacks built-in resource management and requires developers to prevent issues like reentrancy attacks or resource leakage manually \cite{zou2019smart}, Move treats assets as first-class resources. It enforces strict rules through its type system to prevent accidental creation, duplication, or destruction of assets. For instance, when an object is passed by value to a function in Move, it becomes frozen and cannot be reused unless explicitly handled. This mechanism eliminates vulnerabilities such as reentrancy attacks by ensuring that resources are managed safely and predictably.
    \item \textbf{Static Verification and Error Detection}: Solidity often relies on external tools to catch errors and security vulnerabilities, which means issues may only be discovered at runtime \cite{zou2019smart}. In contrast, Move employs a strong static type system and allows for formal verification during compilation. This ensures that potential errors and vulnerabilities are detected early in the development process, reducing the chances of runtime security risks.
\end{enumerate}

Despite the significant advantages offered by the Move language over Solidity, Move remains relatively immature and lacks comprehensive learning resources. The novel paradigms introduced by Move present a learning curve for developers unfamiliar with its structure and conventions, potentially leading to logical errors and, consequently, security vulnerabilities such as access control issues. Therefore, code reviews and safety checks are essential for thoroughly evaluating Move smart contracts to ensure they are free from logical flaws and other security risks. 

Furthermore, Move contracts can still conceal malicious components, including phishing mechanisms, hidden backdoors, and other logic designed to disadvantage general users.

Additionally, in the Move ecosystem, most contracts are not open-source. For example, as of September 2024, over three-quarters of the top Sui Move projects on DefiLLama \cite{SuiDefiL70:online} have not provided their source code. Similarly, almost none of the Coin and NFT projects on Sui \cite{SuiNFTDa91:online, SuiCoinL87:online} have provided verified source code. Even if the code is open-sourced at GitHub, we can't guarantee it is the version that deployed on chain. This lack of transparency makes it extremely difficult for the public to audit deployed smart contracts. Consequently, users are unable to verify the security and reliability of these contracts, hindering their ability to use them with confidence.



\subsection{Decompiler on Move Language}


To address the limited open-source availability of Move contracts in the Sui ecosystem, the Revela Decompiler \cite{verichai85:online} and Move Disassembler \cite{suiexter27:online} were developed. However, despite their potential, both tools have two limitations that discourage users.

\begin{figure}[h!]
    \centering
    \includegraphics[width=1\linewidth]{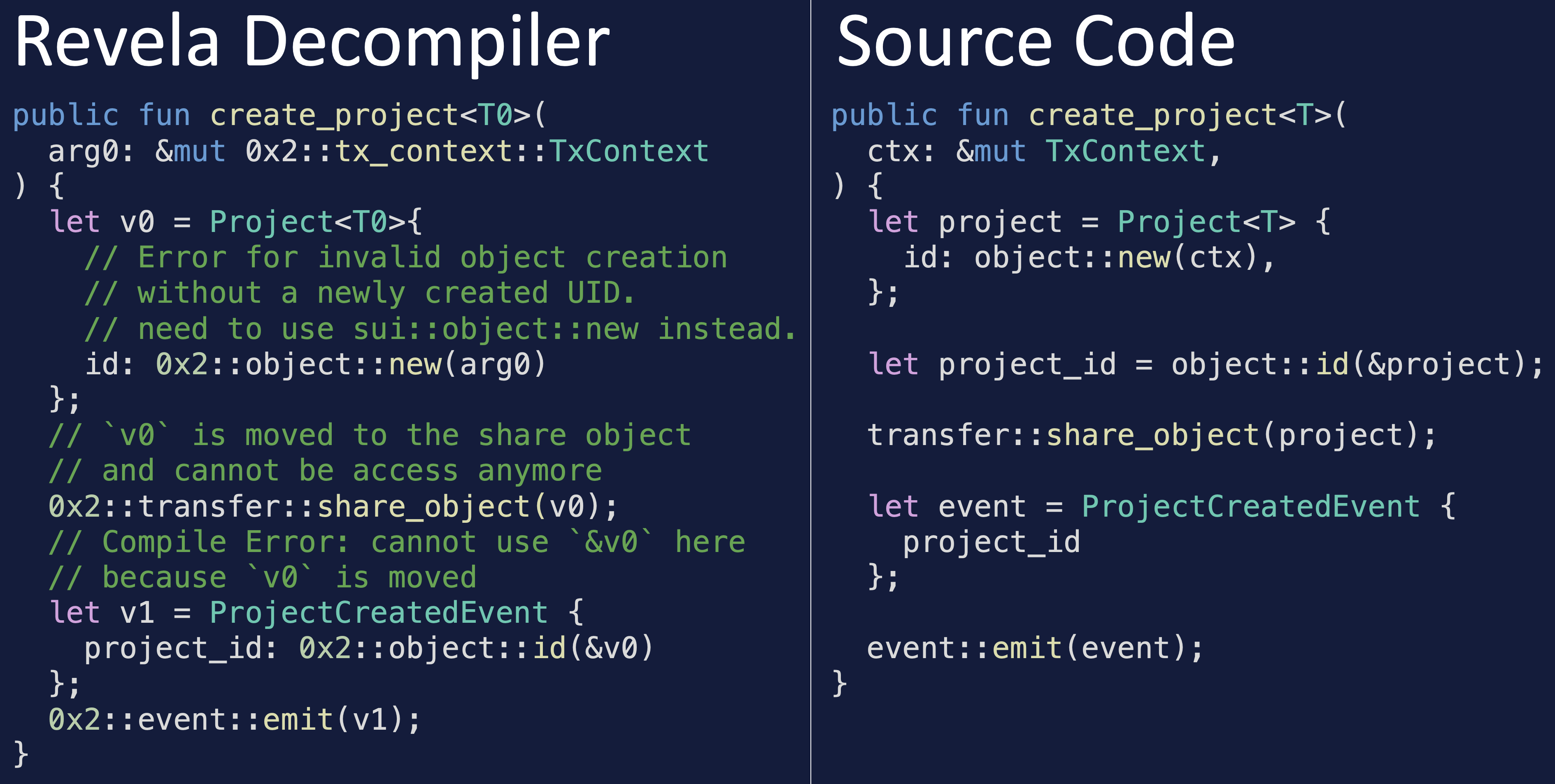}
    \caption{Example illustrated how Revela's output would yield errors when re-compiling with the Sui Move compiler.}
    \label{fig:revela_bug_example}
\end{figure}

Firstly, like other decompilers, they can only generate variable names such as \texttt{v0}, \texttt{v1}, and \texttt{v2}, as illustrated in \autoref{fig:revela_bug_example}. The missing variable names make it difficult for users to perform thorough code reviews on decompiled contracts as they are hard to interpret.

Second, the output from Revela and Move Disassembler cannot be directly recompiled due to the complex rules of the Move compiler. For example, Revela struggles with managing ownership constraints, preventing the access of an object while it is being mutated within the same line, and handling frozen objects, as illustrated in the code comments in \autoref{fig:revela_bug_example}. These issues stem from the Move language's strict resource and ownership rules, which are challenging to decompile accurately. The inability to recompile the decompiled code hinders further verification of security properties and the detection of potential vulnerabilities, such as running unit tests to assess whether the smart contract, when interacting with other deployed contracts, could introduce potential vulnerabilities.

To address this issue, an effective decompiler that allows users to inspect the source code of smart contracts easily and produce re-compilable code is needed. This raises our first research question:

\begin{center}
    \textbf{RQ1:} Can Move bytecode be decompiled into a form that is both human-readable and logically equivalent to source code?
\end{center}

\subsection{AI Augmented Decompiler}

Decompilers can never fully reconstruct the original developer-written code \cite{schulte2018evolving}. Vital elements such as comments, variable names, and types, which significantly contribute to program comprehension \cite{gellenbeck1991investigation, lawrie2006s}, are typically absent from decompiler output.

Recent research has explored the use of artificial intelligence (AI) and machine learning, particularly Large Language Models (LLMs), to augment decompiler output \cite{tan2024llm4decompile, wong2023refining, schulte2018evolving} to make it easier to read and ready to compile. LLMs like OpenAI's GPT-4 have demonstrated remarkable capabilities in code understanding and generation, even with minimal task-specific training data \cite{brown2020language}. For instance, LLM4Decompile \cite{tan2024llm4decompile} fine-tuned LLMs to decompile C language, achieving an 80.49\% re-executability rate on the HumanEval benchmark \cite{chen2021evaluating}. These results indicate that LLMs can enhance the effectiveness and readability of decompilation output by generating both comprehensible and executable code.

However, AI-augmented decompilers have primarily focused on mainstream programming languages like C \cite{tan2024llm4decompile, wong2023refining, schulte2018evolving, chen2021evaluating}, which benefit from abundant data for fine-tuning and evaluation. Their effectiveness hinges on the availability of large labeled datasets, rendering them less applicable to newer, domain-specific smart contract languages like Move, which lack sufficient training data. The limited availability of Move code limits the feasibility of traditional fine-tuning methods that demand extensive datasets. Recognizing these limitations, it is crucial to explore alternative approaches that minimize dependence on large, labeled datasets.

One promising avenue is leveraging the prompt engineering \cite{sahoo2024systematic, schulhoff2024prompt} and few-shot learning \cite{brown2020language} capabilities of LLMs. Due to their extensive pre-training on vast amounts of text data—including code from various programming languages—LLMs can generalize to new tasks with limited examples in the prompt \cite{brown2020language, wang2020generalizing}. This adaptability makes them particularly well-suited for decompiling emerging domain-specific languages like Move. By employing LLMs in this context, we aim to overcome the data scarcity issue and enhance both the readability and executability of decompiled Move code.

It is particularly noteworthy that since the Sui mainnet was launched in May 2023, some LLMs don't have any training data about Sui Move, for example, \textit{GPT-4-1106-preview} only have data until April 2023. This raises the following research question:

\begin{center} \textbf{RQ2}: How stable are LLMs in decompiling Sui Move, and how does their pre-trained knowledge of Move affect their ability to generate re-compilable Move code? \end{center}

By investigating this question, we aim to examine how LLMs' prior knowledge influences their ability to decompile domain-specific languages like Move. Additionally, our research seeks to generalize these findings to other domain-specific smart contract languages, ultimately fostering greater \textbf{transparency} and the ease of \textbf{auditability} of smart contracts across various Web3 ecosystems.

\subsection{Expectancy Theory and Auditing}

Expectancy Theory \cite{vroom1964work, porter1968managerial} is a motivational theory that explains the decision-making process individuals use to pursue certain actions based on the expectation of desired outcomes. The theory suggests that individuals are motivated to act around three key components:

\begin{itemize}
    \item \textbf{Expectancy}: The belief that one's effort will lead to the desired level of performance.
    \item \textbf{Instrumentality}: The belief that achieving the performance will lead to specific outcomes or rewards.
    \item \textbf{Valence}: The value or importance the individual places on the expected reward.
\end{itemize}

In the context of Web3 smart contract auditing, Expectancy refers to the users' belief that they will be able to understand the code effectively. When faced with non-open-source contracts or the limitations of decompilers like Revela, this belief is weakened, reducing their motivation to engage. Instrumentality reflects how users perceive their efforts will result in desirable outcomes, such as detecting vulnerabilities or increasing contract security. Valence, on the other hand, refers to the value users place on these outcomes—how important it is to them to ensure transparency, security, and fairness in smart contracts.

In this paper, we built the SuiGPT Move AI Decompiler (MAD) to address these concerns by generating human-readable and recompilable code by LLMs, increasing users' expectancy of successfully understanding the smart contract logic. As a result, users are more likely to perceive their auditing efforts as instrumental in achieving meaningful, secure results. The higher the Valence users assign to these outcomes—whether it’s protecting their assets or contributing to a more secure ecosystem—the more motivated they are to use tools like MAD to audit contracts thoroughly. To this end, our third research question (RQ3) seeks to explore how Web3 users perceive the outputs and usefulness of the MAD decompiler:

\begin{center} \textbf{RQ3:} How do Web3 users perceive the output of MAD Decompiler, and how do they intend to use it? Will they have any concerns? \end{center}

By investigating RQ3, we aim to understand whether the MAD enhances users' ability to comprehend smart contracts and promotes greater trust and transparency in the Web3 ecosystem.

\section{Development of Move AI Decompiler}
\subsection{Prompt Engineering}
SuiGPT Move AI Decompiler (MAD) leverages the outputs of the Revela Decompiler and the Move Disassembler with prompt engineering techniques to feed into large language models (LLMs), aiming to generate human-readable and re-compilable code.

Intuitively, one might consider feeding the entire output from Revela directly into the LLMs and instructing them to generate the complete code. However, this approach is not feasible in practice. We observed that LLMs struggle to handle long code inputs and often omit parts of the code by summarizing them as comments
or even hallucinations \cite{beutel2023artificial} like omitting or inventing functions. Therefore, it is necessary to process the code in smaller chunks to achieve the desired output.

In our approach, we split the chunk on a per-function basis; then, the input is fed into the LLM using a carefully engineered prompt. We construct our prompt with the following components:

\begin{enumerate}
    \item \textbf{Domain-specific knowledge}: We input knowledge of Sui Move into LLM, including language features, syntax, variable mutation, and object ownership. We also provide instructions on errors commonly encountered in Revela's output, guiding the LLM in fixing these errors during conversion.
    
    \item \textbf{Should and should not instructions}: These instructions emphasize the LLM’s task while avoiding common mistakes, such as ensuring output is well-formatted, using clear variable names, including all necessary type annotations, and should not having hallucinations.
    
    \item \textbf{Few-shot examples}: By providing function code for the input with Revela decompiler, along with the output of the original source code, the LLM is trained to understand the expected input and output format and the syntax of Sui Move. We deliberately selected 17 diverse examples to ensure coverage of the most common scenarios.
\end{enumerate}

Our prompt, after times of iteration, contained 36,120 characters after stringified. Our full prompt is available at \url{https://github.com/EasonC13/MAD_WWW/blob/main/src/prompt/chunk.ts}.

To evaluate the importance of each component in our prompt design, we conducted an ablation analysis with \textit{GPT-4o-2024-11-20} by systematically removing specific sections of the prompt and measuring the recompilation success rate. Using the full prompt, the recompilation success rate reached \textbf{73.33\%}. When ``domain-specific knowledge'' was removed, the success rate dropped to 46.67\%, highlighting the critical role of this component in guiding the model to handle Sui Move-specific syntax and semantics. Removing the ``Should and should not instructions'' or the ``Few-shot examples'' reduced the recompilation success rate to 70\%.

With this well-crafted prompt, we ensured that the model could learn various aspects of the Move language and provided enough examples for it to output well-formed Move code. At the time of development in March 2024, we only had access to GPT-4 with a knowledge cut-off date of April 2023, which did not train on any data about the Move language on Sui. Although we also attempted fine-tuning with \textit{GPT-3.5-turbo}, which is the only model we can fine-tune with at the time, the results were suboptimal, as it was more prone to hallucinations compared to GPT-4.

\subsection{System Development}

We developed both the frontend and backend of SuiGPT MAD using Next.js and Vercel. By utilizing Vercel's edge functions and serverless structures, MAD efficiently processes function chunk decompilation requests in parallel, enhancing overall speed.

Through the MAD web application, users can easily decompile smart contracts on Sui by providing the contract ID. Users can use it with a wallet on Sui to pay the API fee, or they may set the OpenAI API Key in the setting to use it for free. The link to the SuiGPT MAD web application is \url{https://suigpt.tools/mad}, and the source code can be found at \url{https://github.com/EasonC13/MAD_WWW}.

\begin{figure}[h]
    \centering
    \includegraphics[width=1\linewidth]{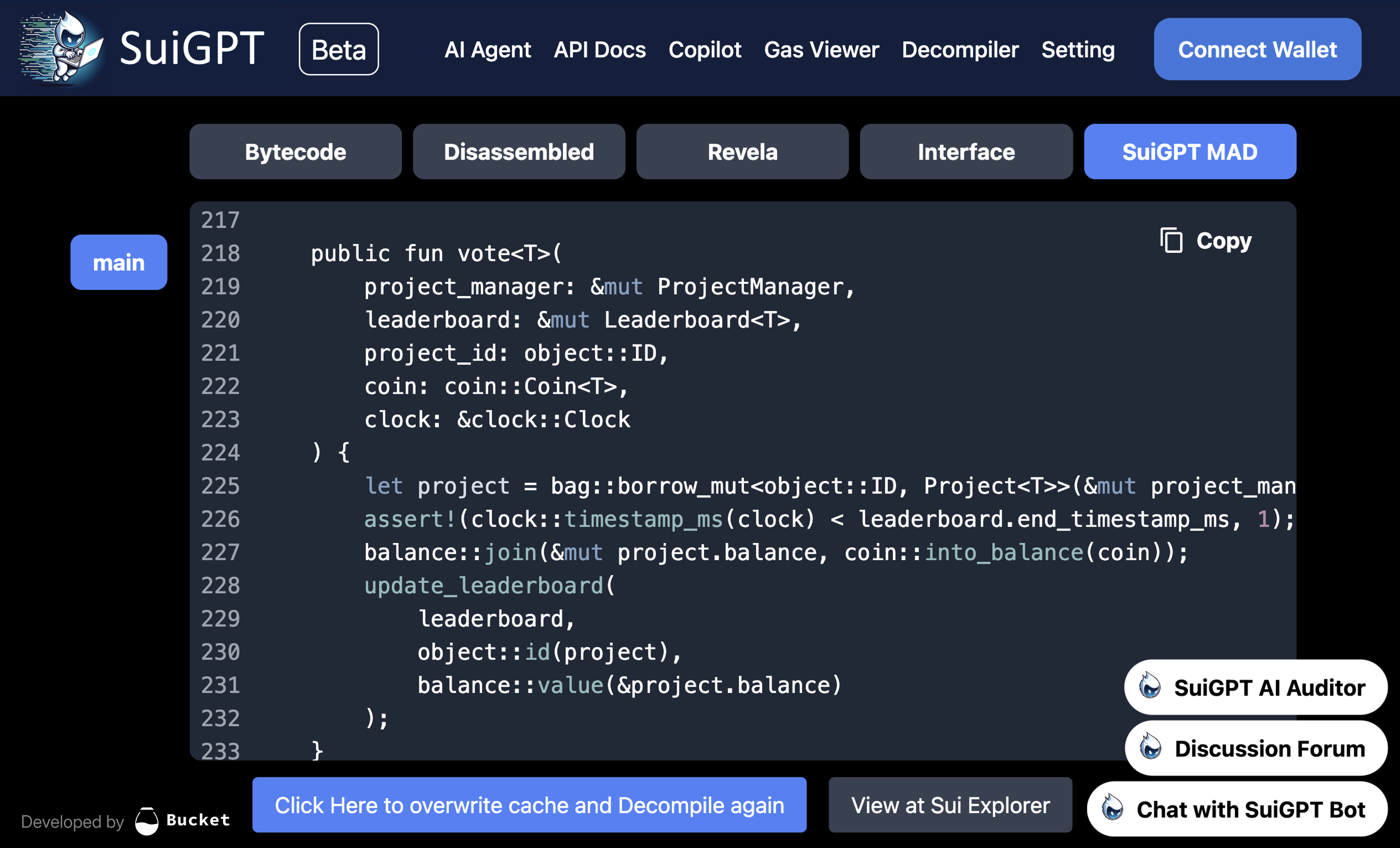}
    \caption{A screenshot of the SuiGPT MAD interface, allowing users to seamlessly explore different versions of decompiled smart contracts, ranging from bytecode and disassembler views to Revela, Interface, and the SuiGPT MAD Decompiler.}
    \label{fig:MAD_interface_screenshot}
\end{figure}

\section{System Evaluation Methods}
To validate the effectiveness of MAD, we created a comprehensive evaluation framework that assesses its performance on both examples with unit tests and real-world smart contracts. This approach tests MAD's ability to handle varying levels of complexity, ensuring reliability in practical applications. The full list of packages we used to evaluate and the evaluation script is available at our GitHub.

When evaluating, the OpenAI API's temperature parameter is set to $0$, and the seed is set to ``123'' to ensure the best reproducibility.

\subsection{Evaluation on Examples with Unit Test}

We first verified that our pipeline could decompile Move bytecodes into logically equivalent source code that can be successfully recompiled, executed, and pass unit tests. We utilized 10 example packages with unit tests from Sui Move's official repositories version 1.22.0
\footnote{We select codes with unit test from \url{https://github.com/MystenLabs/sui/tree/c490f3a19447d17c96cd664729ad39fef32b7230/examples/move}}. These examples provide a known baseline for assessing the functionality and correctness of the decompiled code. We deployed these smart contracts, used MAD to decompile them, and then checked whether the decompiled code could be recompiled and pass the unit tests in the original source code.

\subsection{Evaluation on Real-world Contracts}

To further evaluate MAD in real-world settings, we decompiled the top 30 real-world smart contracts without third-party dependencies, selected from Sui Explorer \footnote{\url{https://suivision.xyz/packages}, data extracted on September 20, 2024}. The exclusion of dependencies was necessary to focus on the contracts themselves, as dependencies add extra complexity to the evaluation pipeline. The contracts chosen represent a diverse set of categories, including gaming, decentralized exchanges and marketplace, Non-Fungible Tokens (NFTs), .etc.

\section{System Evaluation Results}

\subsection{Result on Examples with Unit Test (RQ1)}

The results from our evaluation of smart contracts with unit tests show that 6 out of 10 example contracts decompiled by the MAD Decompiler using \textit{GPT-4o-2024-08-06} were able to be recompiled and passed unit-test successfully without any modification. The remaining 4 were logically correct but encountered some Move language rule check errors inherited from Revela's output, such as using variables after they are frozen. After manually addressing these issues, the unit tests for those contracts also passed \footnote{The decompiled contract for the unit test can be found at \url{https://github.com/EasonC13/MAD_WWW/tree/main/unit_test_gpt-4o_decompiled}}.

These results indicate that MAD is capable of generating code that is logically equivalent to the original source code. However, even though MAD does fix some issues for Move language rules from Revela's output, minor modifications may sometimes be necessary to ensure successful recompilation and execution.

\subsection{Results on Real-world Contracts (RQ2)}

\begin{table}[h]
\centering
\caption{Recompilation Success Rates of Different LLMs}
\begin{tabular}{lcc}
\toprule
\textbf{Model Name} & \textbf{Training Data} & \textbf{Recompilation} \\
                    & \textbf{Cutoff}        & \textbf{Success Rate (\%)} \\
\midrule
\textbf{GPT-4o-2024-11-20}   & \textbf{October 2023} & \textbf{73.3} \\
Gpt-4o-2024-08-06           & October 2023 & 66.7 \\
GPT-4-0125-preview  & December 2023 & 53.3 \\
GPT-4-1106-preview  & April 2023   & 53.3 \\
\bottomrule
\label{table:recompile_rate_of_different_llms}
\end{tabular}
\end{table}

As shown in \autoref{table:recompile_rate_of_different_llms}, during the evaluation of real-world contracts, we found that the model's pre-trained knowledge does not significantly impact its decompilation performance. However, more advanced models tend to perform better. The best performance was achieved by the latest \textit{GPT-4o-2024-11-20}, which had a \textbf{recompilation success rate of 73.3\%}. Although all versions generated logically correct code, errors related to Move language rule checks were observed. Furthermore, we identified some hallucinations. For example, the model incorrectly substituted a custom vector range-checking function with a built-in vector function instead of using the internal function defined in the smart contract. The full output and error of these models can be found in our GitHub \footnote{\url{https://github.com/EasonC13/MAD_WWW/blob/main/evaluation.ipynb}}.

These findings suggest that while MAD enables LLMs to generate logically correct Move code, some syntax errors and hallucinations are still unavoidable, especially in complex real-world contracts. Yet \textbf{the results are good enough to enhance transparency, allowing users to examine the logic of the contracts themselves}. Therefore, we conducted a user study to evaluate this aspect.

\section{User Study and Interview Design}

The capability of MAD to generate decompiled code that is both logically correct and easy to read makes it a valuable tool for users who want to audit non-open-source smart contracts. To further explore its potential, we invited 12 Web3 developers with varying levels of experience in Sui Move language development to participate in our user study. We continued recruiting participants until reaching statistical significance and a saturation point \cite{guest2006many}, where no new insights could be gained from additional interviews.

\subsection{Measure of Workload and Readability}

We used a 7-point Linkert scale for the NASA Task Load Index (NASA-TLX) \cite{hart1988development, hart2006nasa} to measure participants' perceived workload while performing code comprehension and vulnerability detection tasks under various conditions. NASA-TLX is a widely adopted tool for assessing perceived workload across six dimensions: mental demand, physical demand, temporal demand, effort, performance, and frustration. This provides a comprehensive measure of both cognitive and physical strain experienced during a task, offering a deeper insight into task difficulty and workload \cite{hart1988development, hart2006nasa}.

\subsection{90 minutes user study for code reading (P)}
7 participants engaged in a 90-minute session where they worked on a code comprehension task and a 15-minute vulnerability detection session for a smart contract related to a staking and voting campaign \footnote{The code is available at \url{https://github.com/EasonC13/MAD_WWW/blob/main/interview_example/sources/main.move}}. The smart contract was created by an experienced Move developer and included 6 critical vulnerabilities such as access control, function visibility, and backdoor function. These vulnerabilities were inspired by real-world examples of security flaws that have been observed in deployed Sui Move smart contracts. 

Participants were randomly assigned to use either Revela (n = 3) or the MAD (n = 4) for the code comprehension task and vulnerability detection tasks within the same interface; they answered questions such as ``Explain the logic of function X,'' during which their completion times were measured. Following this, they engaged in a 15-minute vulnerability detection task, where the number of vulnerabilities they identified was recorded for comparison. 

After completing the tasks, participants were asked whether they thought the code (MAD or Revela's output) they engaged with was human-written source code or not. Then, they reviewed the correct answers in the vulnerability detection task, code in alternative conditions, and original human-written source code. Then, they participated in a semi-structured interview, during which they filled out the NASA-TLX survey and explored the MAD web applications.

\subsection{30 minutes Quick Interview (Q)}

Additionally, 5 highly experienced Sui Move developers took part in a shorter, 30-minute study. They used the MAD to decompile their own smart contracts, reviewed the decompiled contract code, and compared the output from Revela, MAD, and their own source code. After that, they completed the NASA-TLX survey along with a semi-structured interview. They marked as ``Q'' in the result.

\subsection{Semi-structured interview}

During the completion of NASA-TLX, we engaged participants in discussions to understand the rationale behind their choices. For instance, we asked questions like, "You rated Revela higher than MAD on Mental Demand—could you explain why?"

After completing the NASA-TLX, we asked participants open-ended questions to gain further insights into the potential, limitations, and concerns of MAD. These questions included:

\begin{enumerate}
    \item What are your overall thoughts on MAD?
    \item How and why you might use MAD in the future?
    \item Do you have any concerns about MAD?
    \item How would you feel if the smart contract you deployed was being decompiled by others by MAD?
\end{enumerate}

We applied a thematic analysis \cite{fereday2006demonstrating} approach to analyze the transcripts, focusing on extracting insights from participant responses.

\subsection{Other Informal Conversations}

We also engaged in informal discussions with 6 people from leading Web3 projects on Sui by message or at in-person events to gather their opinions on MAD. In particular, we asked about concerns raised by participants in semi-structured interview sessions.

\section{User Study Results (RQ3)}

\subsection{Perception of Decompiled Code}
When asked, all 3 participants in the Revela condition consider the code is not human-written source code, whereas 75\% (3 out of 4) of participants in the MAD condition believed the code they read was source code.
This suggests that MAD's decompiled code feels like actual source code to the participants.

\subsection{Comparative Analysis of NASA-TLX Scores}

\begin{figure}[h]
    \centering
    \includegraphics[width=1\linewidth]{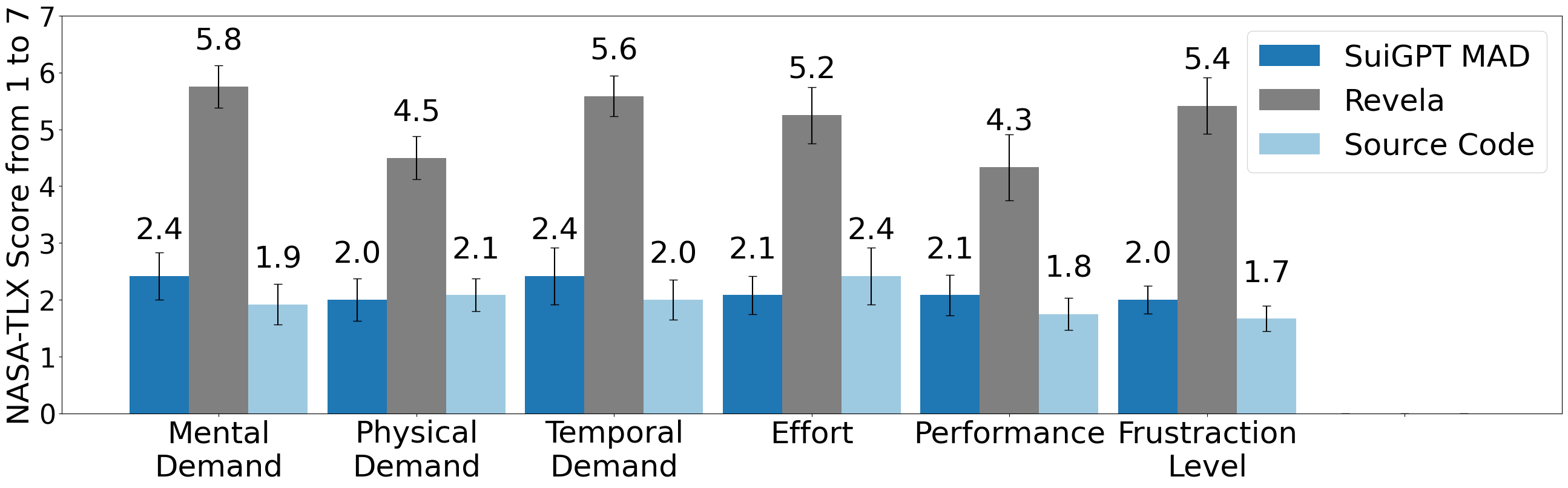}
    \caption{NASA-TLX score results from 1 to 7 for users across MAD, Revela, and Source Code condition. Lower is better.}
    \label{fig:nasa-tlx}
\end{figure}

The result of NASA-TLX across MAD, Revela, and source code is shown at \autoref{fig:nasa-tlx}. Statistical analyses showed that in all dimensions, compared to MAD and source code, \textbf{Revela has a significantly higher workload (all \emph{p} < .01)}. Moreover, there are no significant differences in workload between MAD and the original source code. This suggests that \textbf{MAD provides a reading and auditing workload comparable to the source code}. Due to the page constraints, the detail of the statistical analysis is presented at \autoref{data-analysis-detail-result}.

\subsection{User Interview Result on NASA-TLX}

\subsubsection{Mental, Physical, Temporal Demand and Effort}
\begin{figure}[b]
    \centering
    \includegraphics[width=1\linewidth]{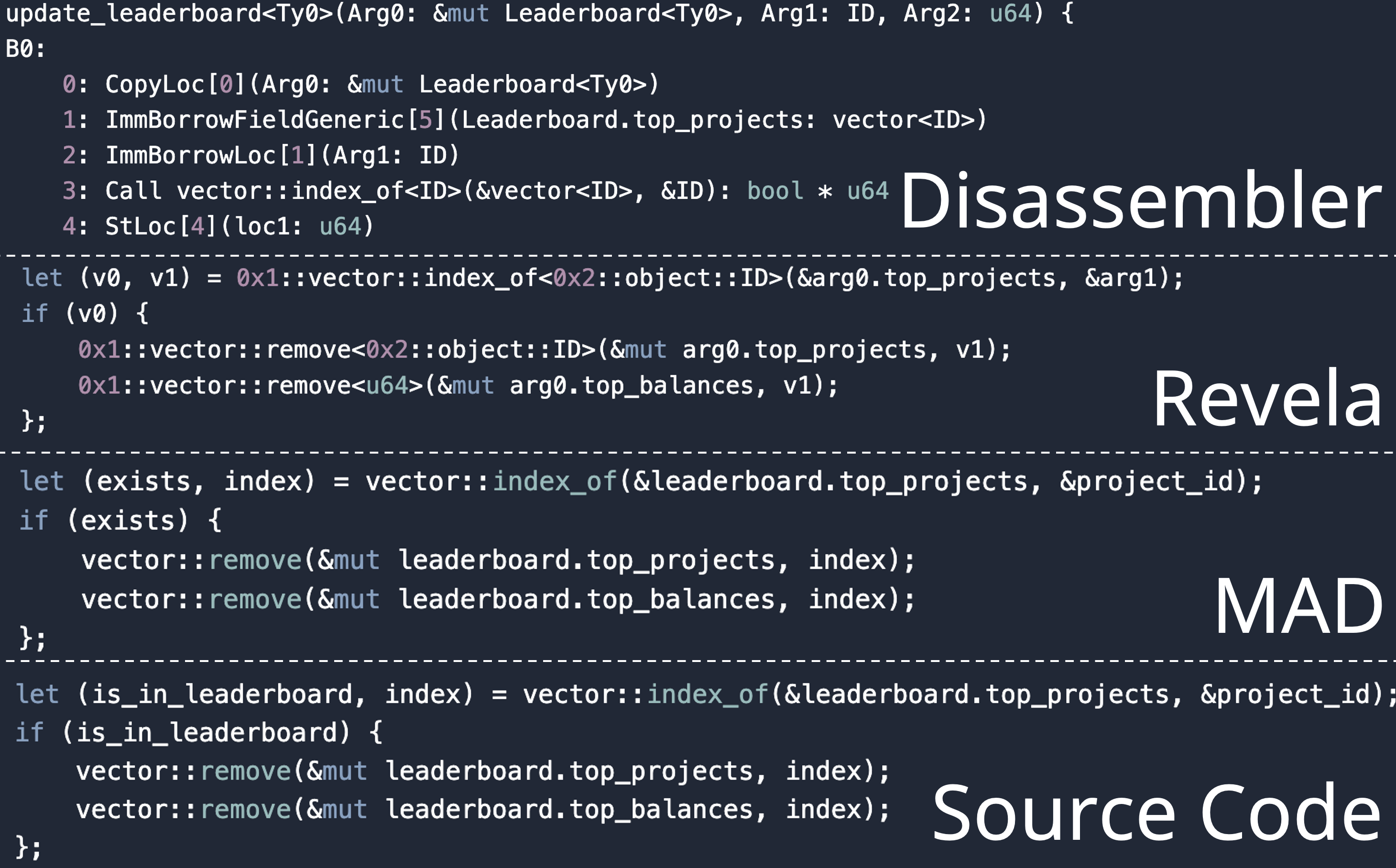}
    \caption{Example output illustrated the difference between Revela, MAD, and Source Code's output.}
    \label{fig:example-code}
\end{figure}

All participants considered that MAD's variable names were easier to understand compared to Revela's, requiring less mental effort. Additionally, because the variable names were more intuitive, there was less need for scrolling around, reducing the physical demand. The ease of comprehension also lowered temporal pressure, making the overall effort lighter. Two participants from the 90-minute group (P) even found MAD's format and variable choices easier to understand than the original source code itself. For example, as illustrated in \autoref{fig:example-code}, one participant noted that MAD's use of the variable name ``\texttt{exist}'' was clearer than the source code's ``\texttt{is\_in\_leaderboard}''.

However, in the quick interview group (Q), participants mentioned that after decompiling their own contracts, MAD's output was not as polished as the original source code. This gap was due to the absence of developer-written comments and advanced Move language features such as Macros \cite{MacroFun16:online}, which define reusable code snippets, and Method Syntax \cite{MethodSy93:online}, which allows functions to be invoked directly from variables. These elements are optimized during compilation but not reflected in the decompiled output. Despite this, participants felt that these differences only lead to a more redundant structure but do not hinder readability or auditing.

These results indicate that MAD can effectively generate smart contract code that closely resembles the original source code, making it readable and suitable for auditing the underlying logic.

\subsubsection{Frustration Level}

Compared to the source code or MAD's code, participants generally felt more frustration when reading code from Revela. The lack of variable names was especially discouraging, making it harder for users to audit smart contract codes.  One participant even admitted during the interview that they almost wanted to give up the experiment while working with the Revela condition. In contrast, participants reading under the MAD condition found it much more intuitive and easy to read, just like reading the original source code.

\subsubsection{Performance}

For the performance dimension in NASA-TLX, all participants perceived their code comprehension and auditing performance to be better in the MAD or Source Code conditions than in the Revela condition. However, for actual performance, we found the number of identified vulnerabilities in the MAD (M = 2.00, SD = 0.82) and Revela (M = 1.33, SD = 1.53) conditions did not differ in a statistically significant manner (t(5) = 0.76, p = .48). Additionally, within a 15-minute time limit, there were 6 vulnerabilities in total, and no participant was able to identify more than half of them.

Additionally, the time spent on function comprehension task showed a similar pattern, with MAD (M = 291.75, SD = 157.86) requiring less time than Revela (M = 450.33, SD = 182.40). However, this difference was not statistically significant (t(5) = 1.24, p = .27), and both conditions took a considerable amount of time.


\subsection{Semi-Structured Interview Results}\label{Semi-Structured Interview Results}

We asked participants open-ended questions to gain further insights into the potential, limitations, and concerns of MAD.

\subsubsection{Usefulness}

All participants found MAD to be a highly useful tool. Many praised its ability to generate decompiled code that closely resembled the original source code, making it easier to read and understand. For example, \textbf{P1} mentioned that ``\textit{MAD's output was quite readable and comparable to source code, which improved their ability to analyze already deployed smart contracts.}'' Similarly, \textbf{Q1} appreciated how MAD made the decompiled code ``\textit{much more readable and helped them better understand the code structure},''.

The ability to analyze non-open-source contracts was a significant advantage noted by participants. \textbf{P4} expressed that ``\textit{MAD would be useful for examining non-open-source projects, meme coins, and decentralized applications to ensure safety before investing.}'' \textbf{P7} highlighted that they could ``\textit{use MAD for auditing purposes, such as decompiling contracts to identify bugs in projects they are considering for investment.}'' \textbf{Q2} pointed out that ``\textit{MAD would be valuable for analyzing specific cases on the network, especially when they have no direct access to the code or its authors.}''

Moreover, MAD was also appreciated for development purposes. \textbf{P6} saw value in ``\textit{using MAD to check other people's code and incorporate relevant sections into their own projects.}'' \textbf{Q5} highlighted that ``\textit{MAD made it easier to understand and debug non-open-source contracts by revealing the internal logic more clearly.}''

Nevertheless, Participants were also concerned about the accuracy of MAD's decompilation. For example, \textbf{P1} mentioned that ``\textit{while he had not encountered major issues yet, they were still cautious about the possibility of hallucinations, such as incorrect or made-up code, which could erode their trust in the tool.}''

\subsubsection{Educational Value}
Several participants emphasized MAD's usefulness in the educational context. \textbf{P2} mentioned that ``\textit{MAD could be a tool for learning purposes, allowing them to learn from well-known production smart contracts and improve their own development skills.}'' \textbf{P3} considered that ``\textit{MAD is helpful for reverse engineering and learning from other developers' code, aiding in debugging and enhancing their own smart contracts.}'' \textbf{Q4} echoed this sentiment, noting that ``\textit{MAD would be highly beneficial for understanding other protocols' code, especially for developers new to Move programming.}''





\subsubsection{Concerns About Their Own Code Being Decompiled}\label{Concerns About Their Own Code Being Decompiled} Participants had mixed feelings about someone else decompiling their own code by MAD. 
Several participants were comfortable with it, as they valued transparency and openness principles of Blockchain. \textbf{P3} expressed no concerns, explaining that ``\textit{if their code was free of vulnerabilities, they had nothing to worry about.}'' \textbf{P4} said ``\textit{they were fine with their code being decompiled, as it aligned with the open nature of blockchain technology.}'' \textbf{Q1} shared a similar view, stating that they ``\textit{preferred open-source practices and saw transparency as key to blockchain's philosophy. }''

Moreover, Some participants, such as \textbf{Q3}, were ``\textit{indifferent to having their code decompiled because their smart contracts were already open-source.}'' \textbf{Q4} even welcomed for being decompiled, mentioning that ``\textit{decompiling their code could make it easier for others to understand, audit, and even learn from it.}''

However, a few participants expressed discomfort with the idea of their code being decompiled. \textbf{P5} admitted they ``\textit{would feel uneasy if their non-open-source code were decompiled, fearing that it could expose their project to risks if vulnerabilities were discovered.}'' \textbf{P6} also raised concerns about ``\textit{competitors potentially cloning their contracts using MAD, which could lead to intellectual property theft.}''

In response to this concern and potential discomfort, \textbf{Q2} and \textbf{Q5} emphasize that ``\textit{all on-chain code is public, meaning determined individuals can reverse engineer it with enough effort.}'' They explained that ``\textit{MAD just streamlines this common process, making code more accessible for analysis, which aligns with blockchain's open nature.}''

In summary, while some participants had concerns about MAD's potential misuse, most were supportive of the transparency it brings to the blockchain space. Many embraced the idea of their code being decompiled as part of maintaining a transparent Web3 ecosystem.

\subsection{Discussion with leading Web3 projects}

We engaged in informal discussions with 6 people from leading Web3 projects on Sui about the concerns raised by our interview participants at \autoref{Concerns About Their Own Code Being Decompiled}. These projects handle millions of dollars through their smart contract protocols on Sui everyday.

It's worth mentioning that during these conversations, one top project founder emphasized how their team had used MAD to accelerate the integration of external smart contracts, particularly when verifying the correctness of complex logic, saving weeks of research time. This demonstrates MAD's potential to simplify integration processes in Web3 projects, offering algorithmic transparency and promoting both technical efficiency and operational benefits.

\subsubsection{Concerns about Exposing Vulnerabilities}

Regarding the concern that MAD might expose vulnerabilities in their smart contracts, people from leading projects generally expressed confidence. Many highlighted that their contracts had undergone professional security audits, which serve as a strong endorsement of the security of their contracts. They pointed out that if a hacker possessed the capability to find vulnerabilities that these audit teams missed, such a hacker would likely be able to reverse-engineer the contract logic regardless of the availability of tools like MAD.

When asked directly about the concerns of people decompiling their contracts, one founder viewed the contracts as effectively open-source once deployed on the blockchain. Their reasoning was that, by being accessible on the blockchain, anyone with enough knowledge could theoretically extract the contract's logic, regardless of MAD's existence. Thus, MAD merely speeds up a process that would otherwise take considerable time and expertise.

\subsubsection{Concerns about Competitors Cloning Contracts}

As for concerns related to competitors potentially decompiling their smart contracts using MAD, people from leading projects indicated that merely cloning the smart contract does not provide a significant advantage. They emphasized that a successful Web3 project relies on much more than its contract alone. Other critical factors include frontend design, backend infrastructure, marketing, and the liquidity value provided to the project. In sum, they highlighted that contracts are just one part of a much larger, complex system that cannot be easily replicated without considerable time and effort.

Regarding intellectual property issues, one manager mentioned that their close-sourced smart contract was protected under a Business Source License (BUSL), meaning that even if others decompiled it, it still has legal protection. Nevertheless, upon further investigation, we found no explicit BUSL declaration with their project. Since the legal issues surrounding the use of decompiled code are outside the scope of this paper, we will not discuss this matter further.

\section{Discussion}


The practical implications of using LLMs to decompile smart contracts extend beyond code readability. By offering a decompilation solution that translates bytecode into logically equivalent and human-readable code, \textbf{MAD makes blockchain auditing more accessible}. This is crucial in the context of blockchain financial and security applications, where trust and transparency are paramount. Our findings indicate that participants have less workload in understanding the code decompiled by MAD compared to existing decompilers like Revela. This suggests that tools like MAD can empower Web3 developers, along with other users, to identify potential risks more easily before engaging with smart contracts.

Moreover, MAD's application in real-world use cases, such as auditing closed-source contracts, introduces a \textbf{powerful tool for preventing exploitation and fraud}. By empowering the developer community to inspect the logic in smart contracts, MAD promotes a more transparent and auditable ecosystem where malicious code is more easily detected. This can help Web3 users avoid vulnerable smart contracts, such as phishing, backdoors, and unfair logic.

In addition to improving transparency, MAD has the potential to be a \textbf{developer education tool}. As noted by several participants, MAD could be used to reverse-engineer non-open-source contracts for learning and development purposes. This educational value offers an opportunity for developers, particularly those new to the Move language, to learn from real-world production codes.

Another key implication is \textbf{the potential generalizability of the MAD methodology to other new domain-specific smart contract languages}, such as Solidity, Rust, and Tact, with the appropriate prompt engineering techniques. This holds true regardless of whether the LLM has been trained with knowledge related to those languages or whether relevant training data is available. This approach would extend AI-powered decompilation, enhancing transparency and auditability across various Web3 ecosystems.

Lastly, \textbf{no participant can identify more than half of the six critical vulnerabilities within 15 minutes task}. However, during the debrief, after viewing the answers, which explicitly pointed out where the vulnerabilities were and how to exploit them, \textbf{all participants validated and understood all six vulnerabilities within three minutes}. This suggests that while MAD's decompiled code is easier to read, users may still need additional support to audit the smart contract. Thus, our future work is enhancing the SuiGPT MAD with features including AI chatbot \cite{chen2023gptutor}, AI auditors with expertly crafted prompts \cite{chen2022decision}, and community forums to help users identify potential vulnerabilities more effectively, as illustrated in \autoref{fig:MAD_interface_screenshot}.

\section{Limitations}

Despite its advantages, MAD still has limitations. Our study identified some minor LLM hallucinations in MAD's decompiled code, such as incorrect function substitutions.
Furthermore, although MAD fixes some errors from the Revela decompiler input related to syntax and resource constraints in Move, manual intervention is still needed to ensure that decompiled code complies with Move's strict compiler rules and can be successfully recompiled.
While these limitations did not hinder code comprehension and auditing, they may limit the tool's effectiveness in complex contracts.

\section{Conclusion}

In this paper, we presented MAD, an AI-powered decompiler aimed at enhancing transparency and auditability in Web3 ecosystems by converting Move bytecode into readable and re-compilable source code. Our evaluations demonstrated MAD’s ability to generate logically correct code, offering significant improvements over existing tools like Revela, especially in enhancing user comprehension and auditability. While there are minor limitations, such as constraints from language rules and occasional inaccuracies that do not impact the logic of the output code, MAD can make non-open-source smart contracts more accessible. This, in turn, contributes to a more secure and auditable decentralized ecosystem.

\begin{acks}
Our study is approved by the Institutional Review Board (IRB) at Carnegie Mellon University.
This research is funded by grants from the Sui Foundation and Carnegie Mellon University Open Source Office Fellowship from the Alfred P. Sloan Foundation.
We extend our heartfelt thanks to the people from Bucket Protocol, Sui Foundation, MystenLabs, Carnegie Mellon University, Sui communities, and paper reviewers for their invaluable feedback and support.
\end{acks}

\bibliographystyle{ACM-Reference-Format}
\balance
\bibliography{reference}

\clearpage
\newpage

\appendix

\section{Data Analysis Detail Results}\label{data-analysis-detail-result}

The Cronbach's Alpha for the NASA-TLX scores across all dimensions was 0.843, indicating good internal consistency.

One-way ANOVA tests and post-hoc Tukey's HSD tests were conducted for each NASA-TLX dimension to determine if there were significant differences among the three conditions. Significant differences were found across all dimensions (\emph{p} < .01). Effect sizes (Cohen's \emph{d} and eta-squared) were calculated to understand the magnitude of differences. The analysis results are presented in \autoref{fig:nasa-tlx}, and \autoref{tab:descriptive-stats}, \ref{tab:anova-results}, \ref{tab:ttest-tukey-effectsize} below.

\begin{table}[ht]
    \centering
    \caption{Descriptive Statistics of NASA-TLX Scores Across Conditions}
    \label{tab:descriptive-stats}
    \footnotesize
    \begin{tabular}{lccc}
        \hline
        \textbf{Dimension} & \textbf{Revela (M ± SD)} & \textbf{Source Code (M ± SD)} & \textbf{SuiGPT (M ± SD)} \\
        \hline
        Mental Demand      & 5.75 ± 1.29             & 1.92 ± 1.24                   & 2.42 ± 1.44             \\
        Physical Demand    & 4.50 ± 1.31             & 2.08 ± 1.00                   & 2.00 ± 1.28             \\
        Temporal Demand    & 5.58 ± 1.24             & 2.00 ± 1.21                   & 2.42 ± 1.73             \\
        Effort             & 5.25 ± 1.71             & 2.42 ± 1.73                   & 2.08 ± 1.16             \\
        Performance        & 4.33 ± 2.02             & 1.75 ± 0.97                   & 2.08 ± 1.24             \\
        Frustration Level  & 5.42 ± 1.73             & 1.67 ± 0.78                   & 2.00 ± 0.85             \\
        \hline
    \end{tabular}
\end{table}

\begin{table}[ht]
    \centering
    \caption{ANOVA Results for NASA-TLX Scores Across Conditions}
    \label{tab:anova-results}
    \begin{tabular}{lccc}
        \hline
        \textbf{Dimension} & \textbf{F (df)} & \textbf{\emph{p}-value} & \textbf{Eta-Squared ($\eta^2$)} \\
        \hline
        Mental Demand      & 29.61 (2, 69)   & < .001                  & 0.6421                          \\
        Physical Demand    & 16.66 (2, 69)   & < .001                  & 0.5025                          \\
        Temporal Demand    & 23.10 (2, 69)   & < .001                  & 0.5833                          \\
        Effort             & 14.97 (2, 69)   & < .001                  & 0.4757                          \\
        Performance        & 10.89 (2, 33)   & < .001                  & 0.3975                          \\
        Frustration Level  & 35.85 (2, 69)   & < .001                  & 0.6848                          \\
        \hline
    \end{tabular}
\end{table}

\begin{table}[b]
    \centering
    \caption{Pairwise T-Test Results, Tukey HSD \emph{p}-values, and Effect Sizes (Cohen's \emph{d}).}
    \footnotesize
    \label{tab:ttest-tukey-effectsize}
    \begin{tabular}{lccccc}
        \hline
        \textbf{Dimension} & \textbf{Comparison} & \textbf{\emph{t}-value} & \textbf{\emph{p}} & \textbf{Tukey HSD \emph{p}} & \textbf{Cohen's \emph{d}} \\
        \hline
        Mental      & Revela vs MAD    & 6.92  & < .001 & < .001 & 1.9964  \\
                          & MAD vs Source Code & 1.73  & .111  & .630   & 0.5000  \\
                          & Revela vs Source Code & 9.93  & < .001 & < .001 & 2.8669  \\
        Physical    & Revela vs MAD    & 5.33  & < .001 & < .001 & 1.5397  \\
                          & MAD vs Source Code & -0.32 & .755  & .984   & -0.0926 \\
                          & Revela vs Source Code & 6.07  & < .001 & < .001 & 1.7525  \\
        Temporal    & Revela vs MAD    & 5.64  & < .001 & < .001 & 1.6271  \\
                          & MAD vs Source Code & 1.60  & .137  & .752   & 0.4628  \\
                          & Revela vs Source Code & 7.66  & < .001 & < .001 & 2.2101  \\
        Effort             & Revela vs MAD    & 6.92  & < .001 & < .001 & 1.9967  \\
                          & MAD vs Source Code & -0.59 & .570  & .860   & -0.1693 \\
                          & Revela vs Source Code & 3.29  & .007  & .0003  & 0.9509  \\
        Performance        & Revela vs MAD    & 4.29  & < .001 & < .001 & 1.2394  \\
                          & MAD vs Source Code & 2.35  & .039  & .846   & 0.6770  \\
                          & Revela vs Source Code & 4.88  & < .001 & .0004  & 1.4102  \\
        Frustration  & Revela vs MAD    & 7.57  & < .001 & < .001 & 2.1842  \\
                          & MAD vs Source Code & 1.48  & .166  & .777   & 0.4282  \\
                          & Revela vs Source Code & 7.36  & < .001 & < .001 & 2.1252  \\
        \hline
    \end{tabular}
\end{table}





\end{document}